\documentclass[pre, twocolumn, superscriptaddress]{revtex4-1}

\usepackage{amsmath}    % need for subequations
\usepackage{graphicx}   % need for figures
\usepackage[colorlinks=true, citecolor=blue, linkcolor=blue]{hyperref}   % use for hypertext links, including those to external documents and URLs

\begin{document}

\title{Absence of Red Structural Color in Photonic Glasses, Bird Feathers and Certain Beetles}

\author{Sofia Magkiriadou}
\affiliation{Department of Physics, Harvard University, 17 Oxford
  Street, Cambridge, MA 02138}

\author{Jin-Gyu Park}
\affiliation{School of Engineering and Applied Sciences, Harvard
  University, 29 Oxford Street, Cambridge, MA 02138}

\author{Young-Seok Kim}
\affiliation{Korea Electronics Technology Institute, 68 Yatap-dong,
  Bundang-gu, Seongnam-Si, Gyeonggi-do, Korea}

\author{Vinothan N.~Manoharan}
\affiliation{School of Engineering and Applied Sciences, Harvard
  University, 29 Oxford Street, Cambridge, MA 02138}
\affiliation{Department of Physics, Harvard University, 17 Oxford
  Street, Cambridge, MA 02138}

\date{\today}

\begin{abstract} 
  Colloidal glasses, bird feathers, and beetle scales can all show
  structural colors arising from short-ranged spatial correlations
  between scattering centers.  Unlike the structural colors arising
  from Bragg diffraction in ordered materials like opals, the colors
  of these photonic glasses are independent of orientation, owing to
  their disordered, isotropic microstructures.  However, there are few
  examples of photonic glasses with angle-independent red colors in
  nature, and colloidal glasses with particle sizes chosen to yield
  structural colors in the red show weak color saturation. Using
  scattering theory, we show that the absence of angle-independent red
  color can be explained by the tendency of individual particles to
  backscatter light more strongly in the blue.  We discuss how the
  backscattering resonances of individual particles arise from
  cavity-like modes, and how they interact with the structural
  resonances to prevent red.  Finally, we use the model to develop
  design rules for colloidal glasses with red, angle-independent
  structural colors.
\end{abstract}

\maketitle

%%%%%%%%%%%%%%%%%%%%%%%%%%  body  %%%%%%%%%%%%%%%%%%%%%%%%%%

\section{Introduction}

Structural color in materials arises from interference of light
scattered from inhomogeneities spaced at scales comparable to optical
wavelengths. Opals and most other familiar examples of structurally
colored materials are ordered~\cite{Sanders1964, Marlow2009}, and as a
result, the color of these photonic
crystals~\cite{MoldingTheFlowOfLight} depends on their orientation
relative to the incident light: they are iridescent. There is another,
less well-studied class of materials with angle-independent structural
colors. These are called \textit{photonic glasses}~\cite{Ballato2000,
  Rojas-Ochoa2004, Garcia2007, Garcia2010}, because the
inhomogeneities form a random, glassy arrangement with short-ranged
order but no long-range order.  As in crystals, the average spacing
between neighboring scatterers in a photonic glass is narrowly
distributed and determines the resonantly scattered
wavelength~\cite{Forster2010, Noh2010AdvMat}. But unlike crystals,
photonic glasses are isotropic, so that the condition for constructive
interference is independent of orientation.  This coloration mechanism
is common in the feathers of birds~\cite{Prum1998, Noh2010AdvMat,
  Saranathan2012}, whose colors are visually indistinguishable from
those of conventional absorbing dyes.  Photonic glasses with
structural colors in the visible have also been produced in a variety
of synthetic colloidal systems~\cite{Takeoka2009, Ueno2009, Ueno2010,
  Forster2010, Harun-Ur-Rashid2010, Lee2010, Kumano2011, Gotoh2012,
  Magkiriadou2012, Takeoka2013}.

However, to our knowledge there are no photonic glasses with saturated
yellow, orange, or red color. While systems with angle-independent
structural red have been reported~\cite{Takeoka2009,
  Harun-Ur-Rashid2010, Kumano2011, Gotoh2012, Park2014}, the color
saturation for long-wavelength hues is poor compared to that for
blue. Interestingly, red angle-independent structural color also
appears to be rare in nature. Birds use structural color only for blue
and green; red colors in bird feathers come from absorbing
pigments~\cite{Saranathan2012}. And while the scales of the longhorn
beetle \textit{Anoplophora graafi} have structural colors that span
the visible spectrum, there are no saturated red colors -- only a pale
purple~\cite{Dong2010}.

\begin{figure}
\centering
\includegraphics{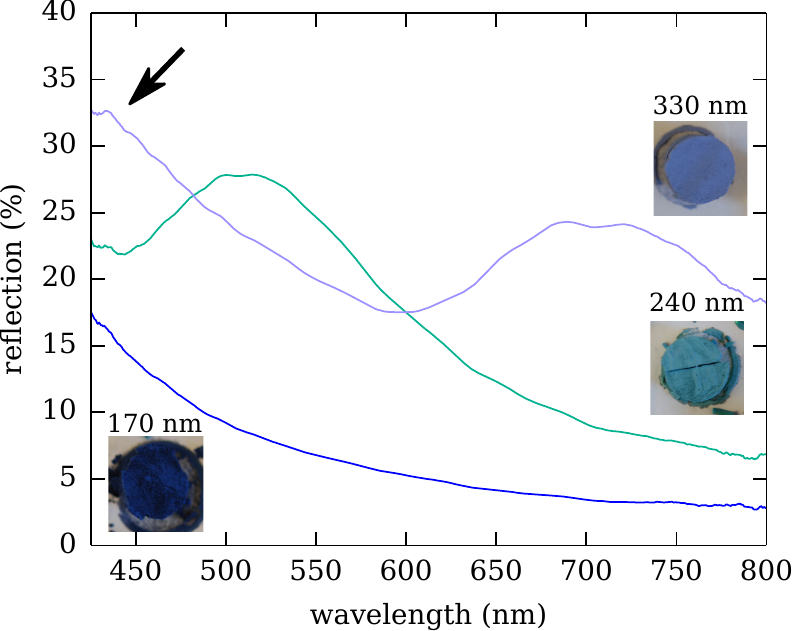}
\caption{\label{fig:data} (Color online) Measured reflectivity spectra
  for three similarly-prepared colloidal glasses of PMMA particles in
  air.  Insets show photographs of the samples and diameters of the
  particles used to make the samples.  The purple sample would appear
  red if not for the high reflectivity in the blue, indicated by the
  arrow.}
\end{figure}

The absence of red photonic glasses does not appear to have been
acknowledged, let alone explained. Previous work on photonic
glasses~\cite{Dong2010, Noh2010AdvMat, Liew2011, Dalba2011,
  Dufresne2009, Forster2010} has focused on structures found in nature
and their biomimetic analogues, nearly all of which are blue. Noh and
coworkers~\cite{Noh2010OptExpress, Noh2010PRE} proposed a theoretical
model based on single and double scattering to explain the optical
properties of these blue samples, and Liew and
coworkers~\cite{Liew2011} and Rojas-Ochoa and
coworkers~\cite{Rojas-Ochoa2004} showed that the inter-scatterer
correlations not only give rise to color, but also suppress multiple
scattering. However, if the color were entirely determined by
correlations, one would expect that all colors could be made simply by
linearly rescaling the structure. As we show below, this approach does
not work (Figure~\ref{fig:data}).

Here we present a model that explains the absence of long-wavelength
structural colors in photonic glasses.  Our model accounts for both
interparticle correlations as well as the scattering behavior of
individual particles. We show that short-wavelength resonances in the
single-particle scattering cross-section near backscattering introduce
a blue peak in the spectrum that changes the hue of a red structural
color to purple. These resonances are not the traditional Mie
resonances, which occur in the total scattering cross-section, but
rather are akin to cavity resonances. The model, which agrees with
measured spectra from synthetic photonic glasses, provides a framework
for understanding the limitations of current photonic glasses and
enables the design of new systems without those limitations.

\section{Experiment}

To demonstrate that resonant scattering from the interparticle
correlations is not sufficient to explain the colors of photonic
glasses, we prepare colloidal glasses from poly(methyl methacrylate)
(PMMA) spheres of various diameters and study their colors with
spectrometry. Figure \ref{fig:data} shows color photographs and
reflection spectra of samples prepared in the same way using three
different particle diameters $d=$ 170, 240, and 330 nm.  These samples
were prepared by mixing one part of an aqueous suspension containing
1\% w/w carbon black (Cabot) and 2\% w/w Pluronic F108 (BASF) with two
parts of a monodisperse suspension of PMMA
particles at 20\% w/w, centrifuging the mixture at 5200 g for 30
minutes, removing the supernatant, and slip-casting on a gypsum
substrate. The final samples are amorphous, dense packings of
particles in air.  The small amount of carbon black suppresses
multiple scattering, making it easier to see the
color~\cite{Forster2010}.

We measure the reflection spectra as a function of wavelength with a
fiber-optic spectrometer (OceanOptics HR2000+) mounted on an optical
microscope (Nikon LV-100).  The samples are illuminated with white
light from a halogen lamp that is collimated by minimizing the
condenser aperture. The divergence angle of the beam is 0.02
radians. The light scattered by the sample is collected by a
50$\times$ objective (Nikon LU Plan Fluor, NA = 0.8) and imaged onto
the detection fiber (OceanOptics QP600-2-UV/VIS). Because the
objective NA is the limiting numerical aperture in our system, our
measurements include light scattered up to a maximum angle
$\theta_{\rm{max}}=\arcsin(0.80)=0.93$ radians. We normalize all
spectra to the reflected intensity from an aluminum mirror, we average
the spectra measured over five different locations on the sample, and
we smooth the spectra using a 50-nm window.

Two peaks are visible in the spectra, one at 515 nm in the $d=$ 240 nm
sample and the other at 710 nm in the $d=$ 330 nm sample.  The ratio
of these peak wavelengths is approximately that of the two particle
sizes $330/240 \approx 710/515$.  The linear scaling is consistent
with the hypothesis that the color is due to constructive interference
of light scattered from correlated regions of particles.  The high
reflectivity of the $d=$ 170 nm sample at short wavelengths suggests a
structural resonance in the ultraviolet (UV), again consistent with
this hypothesis.  Thus the peak wavelength does appear to scale
linearly with the particle size, and hence with the lengthscale of
structural correlations, assuming that the volume fraction and
structure of all of our samples are similar.

However, whereas the $d=$ 170 nm sample appears blue and the $d=$ 240
nm sample appears green, the $d=$ 330 nm sample does not appear
red. The reason is that its reflectivity rises toward the blue,
suggesting a second peak at 430 nm or below.  We have observed similar
spectra in samples made of polystyrene and silica particles whenever
the particle size is 250 nm or larger. Dong and
coworkers~\cite{Dong2010} found a similarly-shaped spectrum, with one
peak in the red and another apparent one at short wavelengths, for
purple longhorn beetle scales~\footnote{see the bottom spectrum in
  Figure 1c of \cite{Dong2010}}. Takeoka and
coworkers~\cite{Takeoka2013} also observed a short-wavelength peak in
the spectrum of their amorphous packings of $d=$ 360 nm silica
particles~\footnote{see Figure 1b, pink curve in their paper}.  In
order to understand the absence of red in all of these photonic
glasses, we must understand the origin of this short-wavelength
reflectivity peak.
 
\section{Theory}

\begin{figure}
\centering
\includegraphics{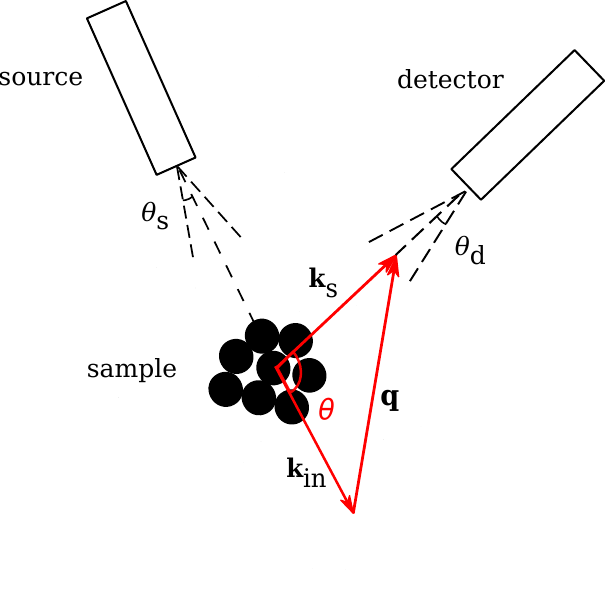}
\caption{\label{fig:geometry} (Color online) Scattering geometry for
  our model.  }
\end{figure}

\begin{figure}
\centering
\includegraphics{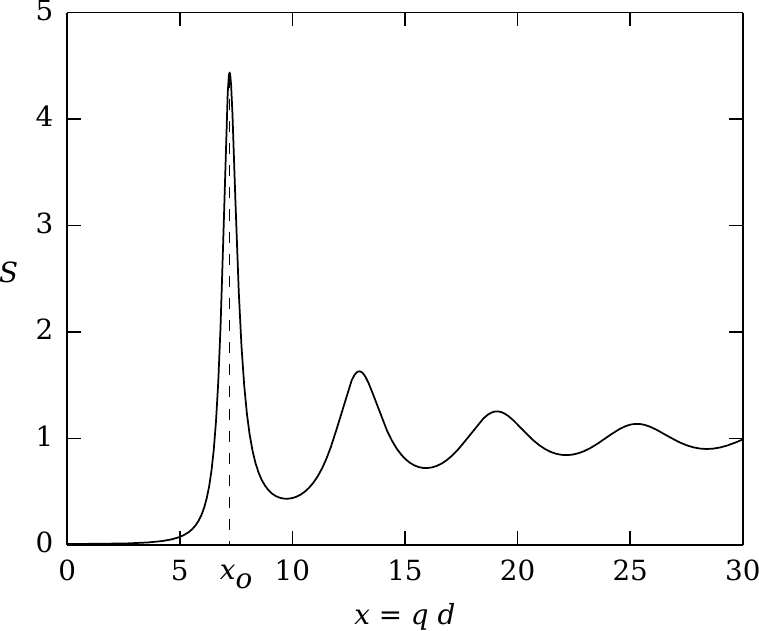}
\caption{\label{fig:structurefactor} Structure factor of a colloidal
  glass with volume fraction $\phi$ = 0.55 calculated from the
  Ornstein-Zernike equation under the Percus-Yevick approximation.}
\end{figure}

To explain our observations, we use a single-scattering model where we
treat scattering from a particle assembly as the result of two
separable processes: scattering from each individual particle, as
described by a form factor, and interference between waves scattered
from the particles, as described by the structure factor of the
glass. This type of single-scattering model has been used to describe the
optical properties of similar systems~\cite{Kaplan1994,
  Rojas-Ochoa2004}. In our analysis, we first use the form and
structure factor to calculate the differential scattering
cross-section of a glass, and then we derive a formula that relates
this cross-section to reflectivity, the quantity that we
measure. According to our model, structural color in photonic glasses
is primarily determined by the peaks of the structure factor. As we
shall show, the form factor can either undermine this color or enhance
it, depending on the structure of the system and the particle
properties.

The scattering geometry that we model mimics a typical experimental
setup, as shown in Figure \ref{fig:geometry}. A colloidal glass is
illuminated by incident light with wavevector
$\mathbf{k}_{\mathrm{in}}$ and scatters light with wavevector
$\mathbf{k}_{\mathrm{s}}$.  The angle between
$\mathbf{k}_{\mathrm{in}}$ and $\mathbf{k}_{\mathrm{s}}$ is $\theta$.
The vector difference $\mathbf{q} = \mathbf{k}_{\mathrm{s}} -
\mathbf{k}_{\mathrm{in}}$ describes the momentum change between the
incident and the scattered wave.  The source can emit light into a
range of angles defined by the source numerical aperture
NA$_{\mathrm{source}} = \sin(\theta_{\mathrm{source}})$, and the
detector can detect light coming from a range of angles defined by the
detector numerical aperture NA$_{\mathrm{detector}} =
\sin(\theta_{\mathrm{detector}})$.

We assume elastic scattering, so that $|\mathbf{k}_{\mathrm{in}}| =
|\mathbf{k}_{\mathrm{s}}| = k = 2\pi n_{\mathrm{eff}} / \lambda$, and 
\begin{equation}
q = 2 k \sin(\theta /2). 
\label{eq:q}
\end{equation}
Here $n_{\mathrm{eff}}$ is the effective refractive index of the medium and
$\lambda$ is the wavelength of light in vacuum.  The effective index
is a weighted average calculated using the Maxwell-Garnett mean-field
approximation~\cite{MaxwellGarnett1904}: 
\begin{equation}
n_{\mathrm{eff}} = n_{\mathrm{med}}\sqrt{\frac{2n_{\mathrm{med}}^{2} +
    n_{\mathrm{p}}^{2} + 2\phi(n_{\mathrm{p}}^{2} -
    n_{\mathrm{med}}^{2})}{2n_{\mathrm{med}}^{2} + n_{\mathrm{p}}^{2}
    -\phi(n_{\mathrm{p}}^{2} - n_{\mathrm{med}}^{2})}}.  
\label{eq:Maxwell-Garnett}
\end{equation}
where $n_{\mathrm{med}}$ is the refractive index of the material
surrounding the particles (air, in our case), $n_{\mathrm{p}}$ is the
refractive index of the particles, and $\phi$ is the volume fraction
occupied by the particles.  We use the effective index because the
particle packings are dense, so that the scattered fields ``see'' an
index intermediate between the particle and medium index.  Although
the Maxwell-Garnett approximation is typically used when the
refractive index variations are much smaller than the wavelength, Vos
and coworkers~\cite{Vos1996} showed that it is a good approximation
even for photonic crystals, and Forster and
coworkers~\cite{Forster2010} have shown the same for photonic glasses.

Assuming perfect monodispersity, we can express the differential
scattering cross-section of a glassy ensemble of particles,
$d\sigma_{\mathrm{glass}} / d\Omega$, as a product of the form factor
$F$ and the structure factor $S$~\cite{Kaplan1994, Rojas-Ochoa2004}:
\begin{equation}
d\sigma_{\mathrm{glass}} / d\Omega = (1/ k^2) F S.
\label{eq:diff_scattering_cross-section_glass}
\end{equation}
The form factor is related to the differential scattering cross
section of a single particle, $d\sigma / d\Omega$, through $F = (1 /
k^2) d\sigma / d\Omega$~\cite{BohrenHuffman}; we calculate $F$ using
Mie theory~\cite{Mie1908, BohrenHuffman}. The structure factor is the
Fourier transform of the pair correlation function of the
particles~\cite{ChaikinLubensky}; we calculate it using a numerical
solution to the Ornstein-Zernike equation with the Percus-Yevick
closure approximation~\cite{PercusYevick1958}.  In so doing, we are
assuming that the structure of our particle glasses is close to that
of a hard-sphere liquid. Figure~\ref{fig:structurefactor} shows $S$
calculated for a volume fraction $\phi = 0.55$ as a function of a
dimensionless wavevector $x = q d$, where $d$ is the diameter of the
particles. We use this structure factor in all our calculations.

In the structure factor, the peak at $x_{o}$ corresponds to the
wavevector $q_{o} = 2 \pi/ d_{\mathrm{avg}}$, where $d_{\mathrm{avg}}$
is the average center-to-center spacing between nearest neighbors.
The peak wavevector $q_{o}$ gives rise to constructive interference and
color, because it sets the relative phase difference between light
waves scattered from neighboring particles.  Resonant scattering
occurs when this phase difference is an integer multiple of $2\pi$.
The wavelengths at which this happens can be determined from
Equation~\ref{eq:q}:
\begin{equation}
\lambda = (4 \pi n_{\mathrm{eff}} d / x_{\mathrm{o}}) \sin(\theta /2). 
\label{eq:resonant_wavelength}
\end{equation}

To describe the intensity of scattered light that reaches the
detector, we integrate the differential scattering cross-section over
the solid angle corresponding to the numerical aperture (NA) of our
detector, taking into account transmission and refraction at the
air-sample interface:
\begin{equation}
\sigma_{\mathrm{detected}} = 1/k^{2} \int_{\phi =
  0}^{2\pi}\int_{\theta_\mathrm{min}}^{\pi} T_{\mathrm{s-a}} F S
\sin(\theta)\, d \theta\, d \phi 
\label{eq:detected_cross-section_glass}
\end{equation}
where $T_{\mathrm{s-a}}$ is the Fresnel coefficient for transmission
at the interface, $\theta_{\mathrm{min}} = \pi -
\arcsin(\mathrm{NA}/n_\mathrm{p})$ is the minimum scattering angle
that we detect, and $\phi$ is the polar angle in the plane
perpendicular to the scattering plane. There is no dependence on
$\phi$ because the particles are spherical and the structure
is isotropic.

To compare our calculations to our measurements we now derive a
relation between $\sigma_{\mathrm{detected}}$ and the measured
reflectivity, $R$. To do this we must account for extinction of light
as it propagates through the sample; the intensity of light scattered
from layers close to the surface is higher than the intensity of light
scattered deeper in the sample because of attenuation by
scattering. Under the assumption of single scattering, this
attenuation scales exponentially with depth following Beer's Law,
$I(x) = I_{\mathrm{o}} e^{-\rho \sigma x}$, where $\rho$ is the number
density of particles, $\sigma$ is the scattering cross-section for the
full solid angle ($0\leq\phi\leq2\pi$, $0\leq\theta\leq\pi$), and $x$
is the distance light has propagated in the
medium~\cite{BohrenHuffman}. If the glass consists of slices of
infinitesimal thickness $\delta x$, the total reflected intensity $I$
is the sum of the intensities $\delta I$ reflected from each slice:
$\delta I = I(x) \sigma_{\mathrm{detected}} \rho\, \delta x$, where
$\sigma_{\mathrm{detected}}$ is given by
Equation~\ref{eq:detected_cross-section_glass}. After integrating both
sides and including the Fresnel coefficients for transmission
($T_{\mathrm{a-s}}$) and reflection ($R_{\mathrm{a-s}}$) at the air-sample
interface, we find
\begin{equation}
R = T_{\mathrm{a-s}} \frac{\sigma_{\mathrm{detected}}}{\sigma} (1 - e^{-\rho \sigma l}) + R_{\mathrm{a-s}}
\label{eq:diffuse_reflection}
\end{equation}
where $l$ is the optical thickness of the sample, or the maximum
distance that light can propagate in it.

\begin{figure}[h]
\centering
\includegraphics[width=8cm]{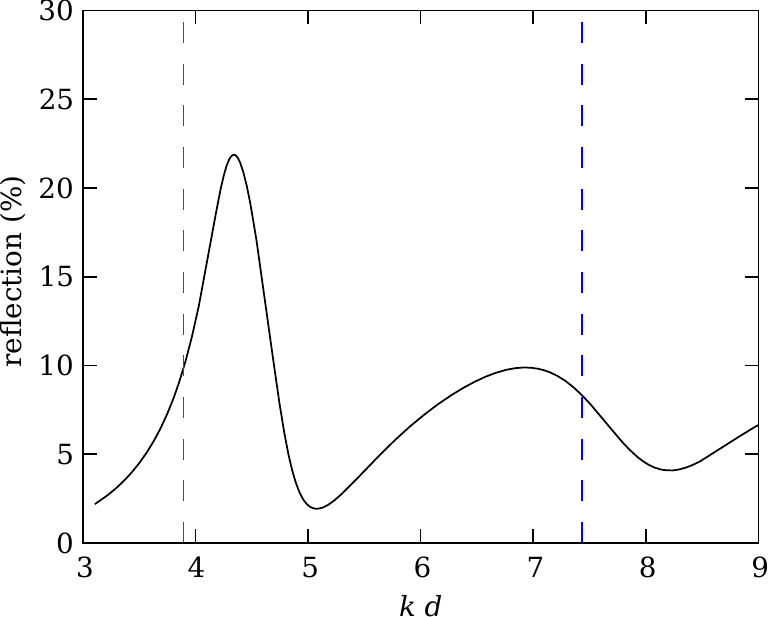}
\caption{\label{fig:rescaled_theory} (Color online) Calculated
  reflectivity as a function of $kd$ for a photonic glass of spheres
  at a volume fraction $\phi=0.55$, as calculated from
  Equation~\ref{eq:diffuse_reflection} with the Fresnel reflection
  coefficient omitted. The vertical dashed lines denote the
  $kd$-values that correspond to the range of visible wavelengths we
  detect, 425 nm (blue line on the right) and 800 nm (red line on the left), when the particle
  size is $d=$ 334 nm.}
\end{figure}

The reflectivity for a glass of PMMA spheres at a volume fraction
$\phi=0.55$, as calculated according to
equation~\ref{eq:diffuse_reflection}, is shown in Figure
\ref{fig:rescaled_theory}. We have omitted the Fresnel coefficient for
reflection at the air-sample interface to better illustrate the
features that arise from scattering from the bulk colloidal glass.
The plot is shown as a function of the scaled particle size $kd$,
where $k = 2 \pi n_\mathrm{p} / \lambda$. All terms in Equation
\ref{eq:diffuse_reflection}---except for the Fresnel reflection
coefficient, which adds an offset in amplitude---scale with $kd$. Thus
this master curve describes reflectivity from a glass made of any
particle size (of the same material and volume fraction), assuming
dispersion is small. Depending on the particle sizes and refractive
indices, different features of the curve will fall within the visible
range; here we mark the edges of our detection range for $d=334$ nm
with the blue (425 nm, right) and red (800 nm, left) vertical dashed
lines.

\section{Results and Discussion}

\begin{figure}
\centering
\includegraphics{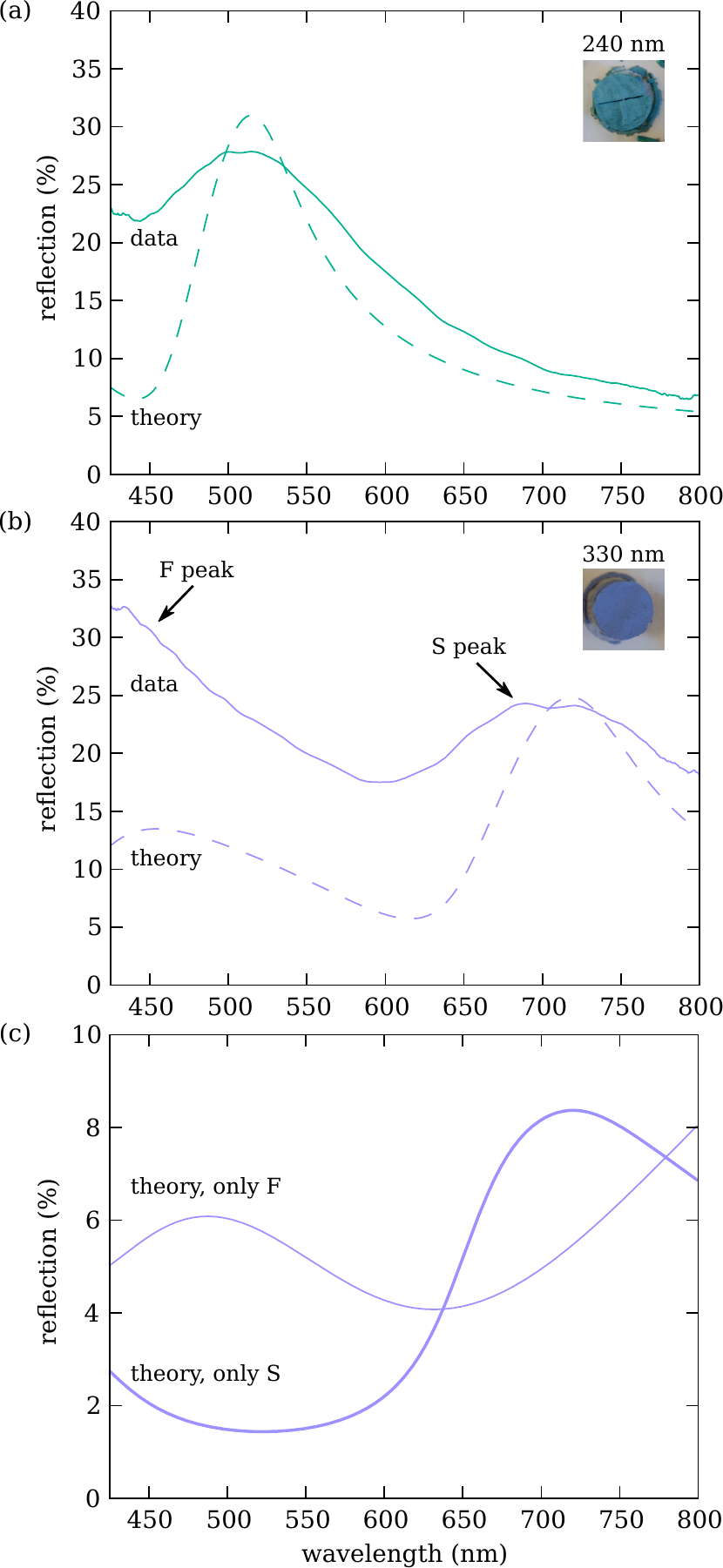}
\caption{\label{fig:data_and_theory_SF_S_F} (Color online) (a),(b)
  Measured (smooth lines) and calculated (dashed lines) reflection
  spectra of colloidal photonic glasses made of PMMA
  spheres. Theoretical spectra are calculated from
  Equation~\eqref{eq:diffuse_reflection} with $\phi = 0.55$ and $l =$
  16 $\mu$m.  Particle diameters that best fit the measured peaks are
  238 nm (a) and 334 nm (b), in good agreement with the measured
  particle diameters. (c) Calculated reflection spectrum for a
  photonic glass made from $d=$ 334 nm particles, including only the
  structure factor (thick curve, divided by 10) and only the form
  factor (thin curve).}
\end{figure}

The calculated reflectivity reproduces the locations of all of the
peaks in our data, as shown in
Figure~\ref{fig:data_and_theory_SF_S_F}. The only free parameters are
the volume fraction $\phi = 0.55$ and the thickness $l=$ 16 $\mu$m.
The peaks predicted by the theory coincide with the peaks in the data
for particle diameters $d=$ 238 nm (Figure
\ref{fig:data_and_theory_SF_S_F}(a)) and $d=$ 334 nm (Figure
\ref{fig:data_and_theory_SF_S_F}(b)); these sizes are in good
agreement with the sizes of the particles measured with scanning
electron microscopy, 240 nm and 330 nm. The calculated reflectivity
for the $d=$ 334 nm system also reproduces the peak in the blue that
we observe in the data.

The model underestimates the amount of light scattered off-resonance,
especially at short wavelengths. We attribute this discrepancy to
multiple scattering.  Since the probability of multiple scattering
increases with the scattering cross-section of individual particles,
its contribution should be more pronounced at short wavelengths,
which is consistent with what we see.

With this model and data at hand we can address our original question:
why does the glass made of 330 nm spheres scatter so much blue light,
when we expect the interparticle correlations to give rise only to a
resonance in the red?  To identify the source of the reflection peaks
in this sample, we compare our data to the reflectivities predicted
from the form factor alone and, in another comparison, the structure
factor alone (Figure \ref{fig:data_and_theory_SF_S_F}(c)). From the
shapes of these two curves we immediately see that the blue peak comes
from the form factor and the red peak from the structure factor,
boosted by another peak in the form factor that occurs at a longer
wavelength.

\begin{figure}
\centering
\includegraphics{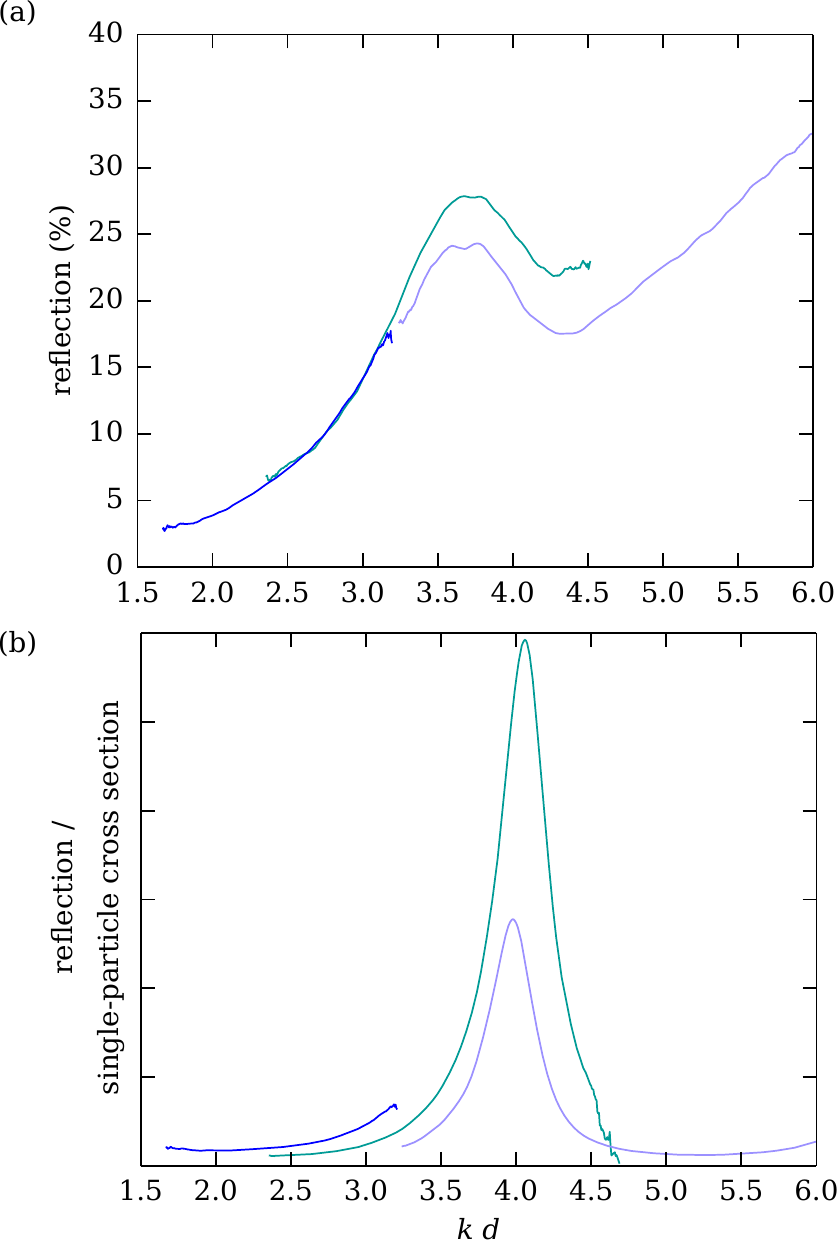}
\caption{\label{fig:all_rescaled_data} (Color online) (a) Reflection
  spectra from Figure~\ref{fig:data} plotted against the dimensionless
  lengthscale $kd$. Note the increased scattering at high $kd$-values
  (short wavelengths) for the $d$ = 330 nm sample. (b) Same as (a) but
  normalized to the single-particle form factor integrated over the
  detected scattering angles, $\sigma_{F, \mathrm{detected}}$, as
  defined in Eq. \ref{eq:detected_cross-section_onlyF}. The increased
  scattering at high $kd$-values disappears, indicating that it is due
  to the form factor. Differences in amplitude of the peaks are likely
  due to differences in the sample thickness $l$.}
\end{figure}

We conclude that the structural colors of our photonic glasses are
determined not only by interference between waves scattered from
correlated particles, but also by resonances in the single-particle
scattering cross-section.  This can also be seen by rescaling the
measured reflectivities by $\sigma_{F,{\mathrm{detected}}}$ and
plotting them as a function of the dimensionless wavevector $kd$
(Figure~\ref{fig:all_rescaled_data}).  Here
$\sigma_{F,{\mathrm{detected}}}$ is the single-particle
differential scattering cross-section, integrated over the angle
subtended by the detector: 
\begin{equation}
\sigma_{F,{\mathrm{detected}}} = 1/k^{2} \int_{\phi =
  0}^{2\pi}\int_{\theta_\mathrm{min}}^{\pi} T_{\mathrm{s-a}} F
\sin(\theta)\, d \theta\, d \phi 
\label{eq:detected_cross-section_onlyF}
\end{equation}
We see that after rescaling the long-wavelength reflectivity peak of
the $d=$ 330 nm sample coincides with the reflectivity peak of the
$d=$ 240 nm sample at the same value of $kd$, showing that these peaks
arise from structural resonances, while the short-wavelength (high
$kd$) peak in the $d=$ 330 nm sample disappears, showing that it
arises from single-particle scattering.

We note that the single-particle resonances responsible for the high
reflectivity at small wavelengths are \emph{not} the so-called ``Mie
resonances.'' Mie resonances are observed in the \emph{total}
single-particle scattering cross-section, obtained by integrating the
differential scattering cross-section over all angles.  The closest
Mie resonance for a 334 nm PMMA particle in our effective medium
occurs at 310 nm.  Instead, as we shall show, the resonances that
contribute to the high reflectivity at small wavelengths occur for
scattering angles near backscattering.

\begin{figure}
\centering
\includegraphics{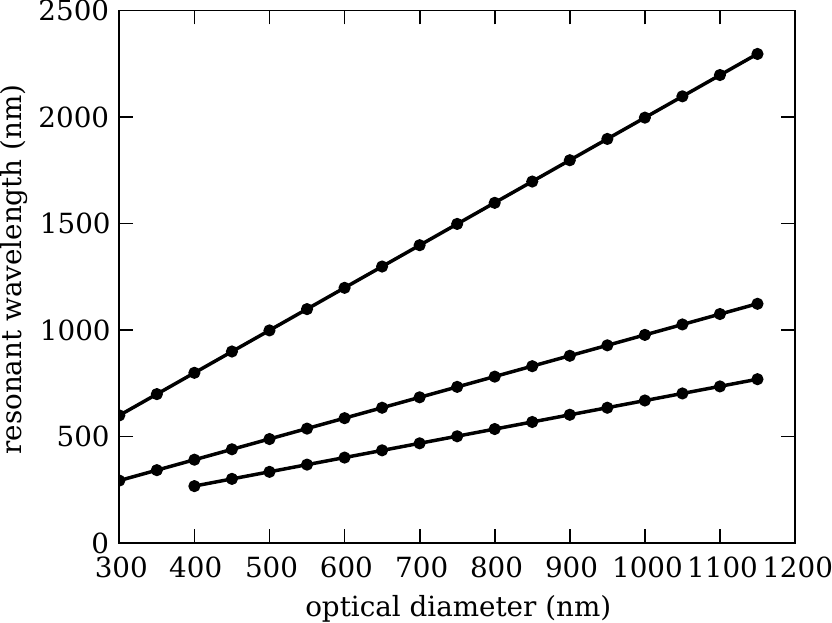}
\caption{\label{fig:resonance_vs_od} Resonant wavelengths of the
  backscattering cross-section, as calculated from Mie theory, as a
  function of optical diameter $n_{\mathrm{p}} d$ for a refractive
  index contrast $m =$ 1.2, which corresponds to that in our
  experimental system. The resonant wavelength follows the linear
  relation $\lambda = 2 n_{\mathrm{p}} d / z$ (solid lines), where $z$
  is the order of the resonance. Lines correspond to different values
  of $z$ (top: $z=1$, middle: $z=2$, bottom: $z=3$).}
\end{figure}

\begin{figure}
\centering
\includegraphics{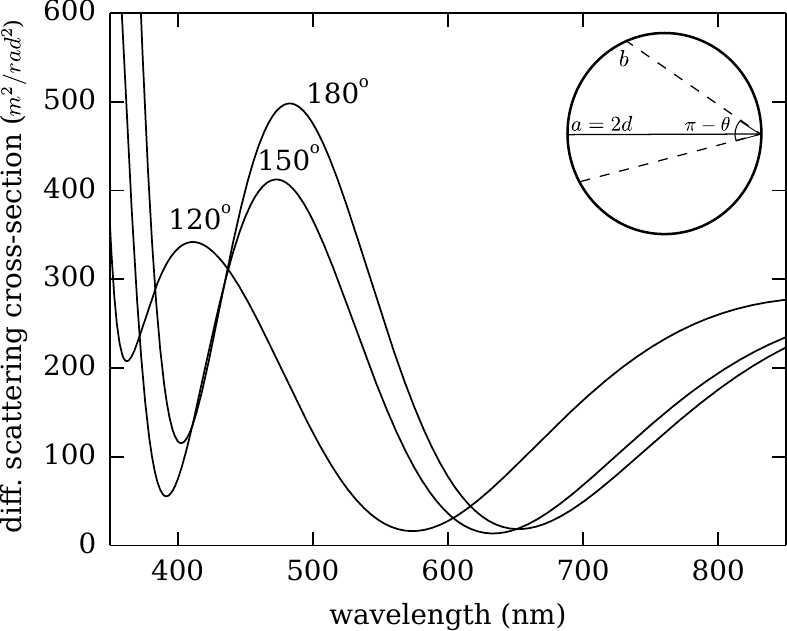}
\caption{\label{fig:differential_cross-sections} Single-particle
  differential scattering cross-sections for various scattering angles
  as a function of wavelength for a 330 nm PMMA sphere in a colloidal
  glass of spheres in air at a volume fraction $\phi = 0.55$. The
  blue-shift in the resonance is consistent with the decrease in
  optical pathlength inside the sphere with decreasing angle: The
  longest possible pathlength $a$ is twice the diameter, and $a > b$
  for any angle that differs from backscattering.}
\end{figure}

We first consider the single-particle scattering cross-section for
pure backscattering.  This is proportional to the differential
scattering cross-section for $\theta = \pi$. As shown in Figure
\ref{fig:resonance_vs_od}, the backscattering cross-section can have
one or more resonances, and the wavelengths at which these occur
increase linearly with the particle ``optical diameter,''
$n_{\mathrm{p}}d$, following the relation
\begin{equation}
\lambda_{\mathrm{resonant}} = 2 n_{\mathrm{p}} d/ z
\label{eq:backscattering_resonances}
\end{equation}
where $z$ is an integer that corresponds to the order of the
resonance. This suggests that these resonances are akin to those of a
Fabry-P\'{e}rot cavity, where constructive interference occurs for
wavelengths that fit an integer number of times within the round-trip
optical pathlength enclosed by the cavity~\cite{Hecht}. Similar
resonances occur in the differential scattering cross-section for all
angles, as shown in Figure \ref{fig:differential_cross-sections}. Our
calculations show that the resonant wavelength shifts toward the blue
as the angle decreases.  This blue-shift is consistent with the
decrease in the round-trip optical pathlength inside the sphere with
decreasing angle (see inset). When the differential scattering
cross-section is integrated over the detected solid angle ($\theta =
150^\circ - 180^\circ$ for our detection numerical aperture of 0.8,
after refraction), these resonances, though broadened, persist (Figure
\ref{fig:data_and_theory_SF_S_F}(c)).  We therefore conclude that
constructive interference of light inside the particles contributes to
enhanced reflection at short wavelengths.

Our results show that the absence of red structural color in photonic
glasses can be attributed to the blue scattering resonances within the
component particles. These resonances occur in addition to the
resonances from interparticle correlations, meaning that structural
color in photonic glasses arises from the combination of resonant
scattering from the structure \textit{and} from the individual
particles.  For blue and green structural colors, these
single-particle scattering resonances do not affect our perception of
color, because they occur in the UV.  While these observations are
based on glasses of PMMA spheres, the same principles apply to most
typical colloidal materials, such as polystyrene and silica, whose
refractive indices are not significantly different from that of that
of PMMA.  The chitin particles in beetle scales have similar optical
properties~\cite{Dong2010}.

Although the single-particle resonances we observe should also occur
in photonic crystals, the structure factor in crystals is much more
sharply peaked than in glasses, so that structural resonances can
dominate the single-particle resonances, and photonic crystals can
have structural red color. The price to be paid for this red color is
strong angular dependence: changing the angle between source and
detector changes the resonant wavelength, leading to iridescence.
Photonic glasses have a lower and broader peak in the structure factor
as a consequence of their disorder.  The breadth of the peak allows
the glass to maintain nearly the same structural color over a range of
angles between the source and the detector.  At the same time, the low
peak amplitude (relative to a crystal) makes the color of the glass
susceptible to contamination from the single-particle resonances.
Thus we see that there is an inherent tradeoff between angular
independence and structural red color.

Can this tradeoff be broken?  One obvious way is to introduce a dye
that absorbs blue light. However, this is how traditional colors are
produced; red paint, for example, consists of strongly scattering
particles mixed with pigment particles that absorb the incident and
scattered blue light.  In contrast, structural color must arise from
resonances that allow scattering at certain wavelengths to dominate
scattering at all other wavelengths. The absence of
wavelength-dependent absorbers makes structural color appealing for
applications such as coatings, because all colors can be produced from
the same materials.  Therefore we pursue a different way to break the
tradeoff between angular independence and long-wavelength structural
color.

To create a red structural color, we would need to manipulate the
resonances from two independent processes: single-particle scattering
and coherent scattering from the particle assembly.  The
characteristic scale for the single-particle scattering resonances is
the particle optical diameter $n_{\mathrm{p}}d$
(Equation~\ref{eq:backscattering_resonances}), and for the structural
resonances it is the effective interparticle spacing
$n_{\mathrm{eff}}d_{\mathrm{avg}}$
(Equation~\ref{eq:resonant_wavelength}).  Our control parameters are
therefore the particle diameter $d$, its refractive index
$n_{\mathrm{p}}$, the interparticle spacing $d_{\mathrm{avg}}$, and
the index of the medium $n_\mathrm{med}$, which determines the
effective index $n_\mathrm{eff}$.

Red photonic glasses could be made by tuning these control parameters
to blue-shift the second resonance of the form factor into the UV
while keeping the peak of the structure factor at long
wavelengths. This can be achieved with particles that have a small
optical diameter, as shown in Figure \ref{fig:resonance_vs_od}.  The
simplest way to reduce the optical diameter is to use particles with a
refractive index smaller than that of the medium.  In these
\textit{inverse} glasses, the diameter of the particles is about the
same as the spacing between their centers ($d_{\mathrm{avg}} \sim
\sqrt{6} d/3$), but their lower refractive index makes the optical
pathlength inside each particle shorter than the optical pathlength
between two particles. As a result, the form factor resonances are
blue-shifted compared to those of our PMMA colloids of similar size,
while the structure factor resonances remain at about the same
wavelength.

Another effective way to blue-shift the form factor resonance while
not shifting the structural resonance is to decouple the particle size
from the interparticle spacing and to use smaller particles as the
scatterers. This can be done by packing core-shell particles with a
strongly scattering core and a transparent shell. We have already
demonstrated that this approach enables the creation of full-spectrum,
angle-independent structural pigments~\cite{Park2014}.  Such pigments
still suffer from poor color saturation in the red compared to the
blue, but the short-wavelength reflectivity is substantially reduced
relative to colloidal glasses made of homogeneous (that is, not
core-shell) particles.

\begin{figure}
\centering
\includegraphics{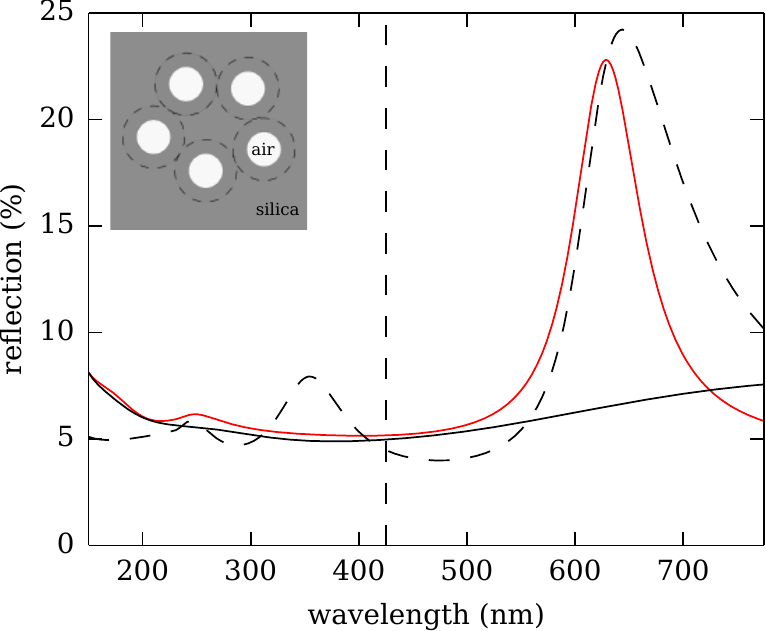}
\caption{\label{fig:reflection_red} (Color online) Calculated
  reflection for an inverse glass of core-shell spheres engineered to
  scatter most strongly in the red. The continuous red curve with the
  peak at 632 nm includes both the structure and the form factor, the
  dashed black curve only the structure factor (divided by 2), and the
  black continuous curve only the form factor.  The vertical dashed
  line marks 425 nm, the low-wavelength limit in our spectral
  measurements. The optimal design has air cores with diameter 260 nm
  and silica shells with diameter 280 nm, and the particles are
  embedded in a silica matrix. The reflectivity peak at 632 nm is
  primarily due to the structure factor and determined by the shell
  diameter. One form-factor resonance occurs in the near-IR, boosting
  the structural resonance in the red, and another occurs deep in the
  UV (at about 150 nm), too far away from the visible regime to affect
  the color.}
\end{figure}

One can combine the core-shell and inverse-structure approaches to
design a system with a single visible resonance at long wavelengths.
In particular, the shell diameter could be chosen such that the
interparticle spacing gives rise to a resonance in the red, and the
core diameter chosen such that the first-order peak in the form-factor
boosts the peak from the structure factor, while the second-order
form-factor peak is fully in the ultraviolet.  Our calculations show
that such a structure could be made from core-shell particles with air
cores and silica shells, embedded in a silica matrix, as shown in
Figure~\ref{fig:reflection_red}.

Curiously, the photonic structures found in bird feathers resemble
these inverse glasses: they often consist of air pockets in a matrix
of $\beta$-keratin~\cite{Noh2010AdvMat, Saranathan2012} that has a
refractive index close to that of
silica~\cite{Leertouwer2011}. However, birds do not seem to have taken
advantage of their inverse structures for colors other than blue.
They rely instead on pigments to acquire yellow, orange and red
colors~\cite{Saranathan2012, Stoddard2011}. Whether this is due to a
physical effect not accounted for in our model or is the result of
evolution or chance remains to be seen.

\section{Conclusions}

We have described a physical mechanism that explains the scarcity of
angle-independent structural red color in photonic glasses. Our model
shows that interparticle correlations alone are not sufficient to
understand the reflectivity of photonic glasses. The scattering
behavior of their constituent particles plays an equally
important---and previously unrecognized---role. In particular,
interference of light \textit{inside} the particles can lead to
enhanced scattering at wavelengths other than those related to the
interparticle correlations. Our model describes our experimental
observations well, and it can be used to guide the design of new
photonic glasses with purely red structural color.

To this end, we have shown that it is possible to control the
locations of both the single-particle scattering resonances and the
structural resonances by tuning the refractive indices, particle
sizes, and interparticle distances. Our model predicts that inverse
glasses made of core-shell particles with a low-index core and a
high-index shell that is index-matched to the medium might produce
angle-independent structural red color. If future experiments show
that such structures show poor color saturation in the red, these
results would suggest that another mechanism, such as multiple
scattering, should be accounted for.  If successful, these structures
would complete the palette of visible colors achievable with photonic
glasses, opening the path to their use in practical applications such
as long-lasting dyes or reflective displays.

\begin{acknowledgments}
We thank Ben Rogers, Max Lavrentovich, Jason Forster and Eric Dufresne
for helpful discussions and Rodrigo Guerra for the PMMA
particles. This work was supported by an International Collaboration
grant (No.Sunjin-2010-002) from the Korean Ministry of Knowledge
Economy and by the Harvard MRSEC (NSF grant no. DMR-0820484). It was
performed in part at the Center for Nanoscale Systems (CNS) at Harvard
University, which is supported by the National Science Foundation
(ECS-0335765).
\end{acknowledgments}

%\bibliography{manuscript}
%

\end{document}